**Adam SIEMIEŃSKI**, Department of Biomechanics, University School of Physical Education in Wrocław

# CAN ALL MUSCULAR LOAD SHARING PATTERNS BE REGARDED AS OPTIMAL IN SOME SENSE?

Summary: Muscles crossing a joint usually outnumber its degrees of freedom, which renders the motor system underdetermined. Typically, optimization laws are postulated to cope with this redundancy. A natural question then arises whether all muscular load sharing patterns can be regarded as results of optimization. To answer it we propose a method of constructing an objective function whose minimization yields a given load sharing pattern. We give necessary conditions for this construction to be feasible and investigate its uniqueness. For linear load sharing patterns the Crowninshield and Brand (1981) objective function is reproduced – nonlinear ones require a more general formula.

1. INTRODUCTION

In order to answer the title question, we need to revisit the rationale of using optimization theory to solve the muscle load sharing problem. It is based on three premises. First, muscles actuating a joint usually outnumber its degrees of freedom and consequently a desired joint torque may be produced by many different muscle activation patterns. Second, despite this redundancy muscle activations during well learned motor tasks are highly stereotypical, which suggests the existence of a law justifying the selection of one activation pattern out of infinitely many feasible ones. And third, ideas from evolutionary biology help interpret this law as a minimum principle.

Within such a formulation, the set of muscle activation levels actually selected by the central nervous system for a given task is such that the desired external joint torques are produced and a quantity depending on muscle activations attains a minimum. In terms of optimization theory this becomes a problem of finding a constrained minimum of a cost function. While there is a general agreement as to the form of the constraints – they should express the torque equilibrium conditions at joints – the proper choice of the cost function is less obvious. In fact, the whole three-decade-long history of efforts to solve the force sharing problem could be told by enumerating the cost functions hypothesized so far [2]. As different types of cost functions usually lead to different force sharing patterns between muscles, a natural question arises whether an appropriate cost function could be found for any observed force sharing. Such questions fit within the inverse optimization approach where variables are sought that are optimized by the observed patterns of behavior, i.e. instead of seeking the best possible solution to a problem, one asks: if this structure or pattern of movement is the best possible solution to a problem, what was the problem? [2]

However, the current approach to the force sharing problem interpreted in terms of inverse optimization seems little more than just another example of the trial and error method. Indeed, cost functions are guessed/hypothesized and then an optimization procedure is performed to check its outcome against the observed force sharing patterns [5]. Although there are

Can all muscular load sharing patterns be regarded as optimal in some sense?

examples, outside of biomechanics, of a more systematic attitude to inverse optimization which goes beyond trial and error and reduces it to an optimization problem of finding the best parameter adjustment in an assumed form of objective function [1], they can hardly be regarded as direct solutions of the problem.

Now turning back to the original question, we will try to answer it by proposing a *direct* method of constructing an objective function whose minimization results in a given load sharing pattern, and by deducing its feasibility conditions.

2. METHOD

In order to explain the main idea of the construction let us first formulate a force sharing optimization problem and its inverse optimization counterpart. Let us imagine a one-degree-of-freedom joint spanned by $N$ muscles. The torque equilibrium equation reads

$$\sum_{i=1}^{N} r_i \cdot F_i = M \qquad (1)$$

where r, F, M denote lever arm, muscle force and joint torque, respectively. Now the force sharing problem consists in finding such muscle forces $F_1, F_2, F_3, \ldots F_N$ that Eq. 1. is satisfied and an objective function

$$K = K(F_1, F_2, F_3, \ldots) \qquad (2)$$

is minimized. Mathematically, this is a typical constrained minimization problem, which can be solved using the Lagrange multiplier method. This means that given an objective function (2) and the constraint (1) one can derive force-force functions, which define force sharing patterns. Some additional assumptions are usually made as to the form of the objective function to the effect that it becomes additive and homogeneous

$$K = \sum_{i=1}^{N} a_i \cdot g(F_i) \qquad (3)$$

with $g$ representing a monotone function. Eq. 3. reflects the idea that the minimized quantity should be extensive (as for example energy) and scalable (big muscles are just bigger, not otherwise different). Note that this *is* a restriction and not all objective functions can be expressed as in Eq. 3.: e.g. sum of muscle forces, sum of muscle forces squared, sum of muscle forces cubed are of this general form but some other less mundane characteristics as effort or fatigue – are not.

The associated inverse optimization problem would consist in finding the objective function based on known constraints and force sharing patterns. Let us consider the more modest task of finding the form of function $g$ as opposed to the more ambitious problem of identifying the function of many variables $K$. Our inverse optimization problem will then consist in finding $g$ based on known constraint (1) and known force-force function $h$

$$F_j = h(F_i) \qquad (4)$$

for at least one pair of muscles, $F_i, F_j$, engaged in the joint.

Can all muscular load sharing patterns be regarded as optimal in some sense?

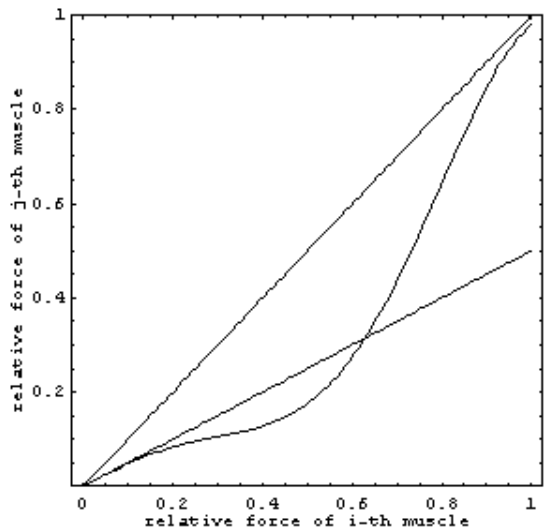

Fig. 1. An example of force-force function *h* shown against identity line and a straight line whose slope is equal to *h'(0)*

An exemplary diagram of function *h* for relative muscle forces is shown in Fig. 1. together with two straight lines – an identity line and a straight line whose slope is equal to *h'(0)*.

Reformulating the problem in terms of Lagrange multipliers will help establish the relationship between function *g* and function *h*.

$$\sum_{i=1}^{N} a_i \cdot g(F_i) + \lambda \cdot (M - \sum_{i=1}^{N} r_i \cdot F_i) \rightarrow Minimum \qquad (5)$$

Differentiating (5) with respect to *Fi* yields

$$a_i \cdot g'(F_i) - \lambda \cdot r_i = 0 \qquad (6)$$

for every *i*. For the two muscle forces of Eq. 4. one obtains therefore

$$g'(F_j) = \frac{r_j a_i}{r_i a_j} g'(F_i) \qquad (7)$$

Central to the method being presented here is the observation that after suitable substitutions Eq. 7. becomes Schröder's functional equation [4] for the unknown function *f*, i.e. the derivative of the objective function *g*, containing the known function *h*

$$f(h(x)) = s \cdot f(x) \qquad (8)$$

Solving the inverse optimization problem is therefore equivalent to solving this equation, and then integrating its solution.

Can all muscular load sharing patterns be regarded as optimal in some sense?

3. RESULTS AND CONCLUSIONS

Schröder's functional equation was not solved by Schröder himself but by Koenigs [4] who formulated a general theorem stating that if a function of real or complex variable $h$ has an attractive fixed point at 0, i.e. , $h(0)=0$, $0<h'(0)<1$, then a general solution of Eq. 8. is given by the following formula

$$f(x) = C \cdot [\underset{n \to \infty}{Lim} \frac{h_n(x)}{h'(0)^n}]^p \qquad (9)$$

where $h_n$ denotes $n$-th iteration of $h$, and $p$ depends on $s$ (Fig. 2.). Eq. 9. defines $f$ up to one multiplicative constant $C$. The function $g$ can then be obtained by integration, giving rise to another (additive) constant. The objective function K (Eq. 3.) can thus be constructed for virtually any differentiable monotone load sharing pattern $h$.

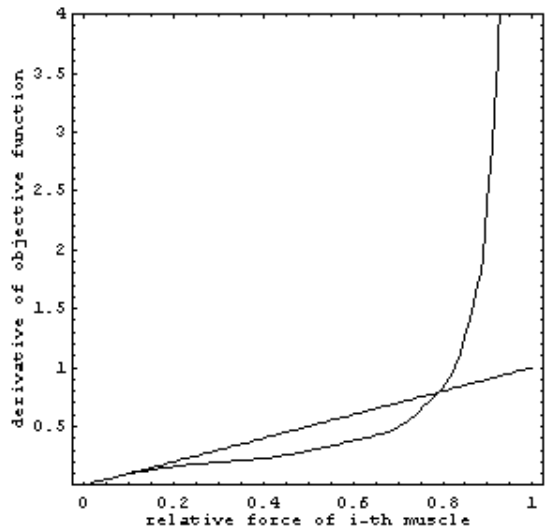

Fig. 2. Koenigs solution of Schröder's equation, i.e. function $f$, corresponding to force-force function $h$ of Fig.1., shown against identity line

Moreover, within this framework, load sharing patterns for other pairs of muscles crossing the same joint turn out to be fractional iterates of $h$. This conclusion may seem surprising as it means that within the optimization approach the force-force functions for all pairs of muscles crossing the joint are related and can be easily calculated one from another. It obviously becomes less strange when one realizes the surprising fact that thanks to Koenigs theorem the objective function behind all the observed force sharing patterns is constructed based on just one of them.

In a special case of linear $h$ the objective function proposed in [3] is reproduced, and for such an objective function the two above mentioned facts become obvious. The above limit (Eq. 9.) can also be evaluated exactly for some nonlinear functions $h$, but for a general force-force function, e.g. one obtained experimentally, an approximate formula must be sought instead by stopping iteration at some $n$. Practically, due to quick convergence, stopping at $n=10$ usually results in $f$ indistinguishable from the limit. Fig. 3. illustrates just how quickly functions defined by Eq. 9. converge to the limit as $n$ increases. Six curves are shown there but the two corresponding to $n$ equal to 9 and 100 are indistinguishable. This effect is also

Can all muscular load sharing patterns be regarded as optimal in some sense?

illustrated in Fig. 4. where the ratio $f(h(x))/f(x)$ is shown as a function of $x$, the relative muscle force. The same values of $n$ are used as before and now six distinct curves are visible but the two corresponding to $n$ equal to 9 and 100 differ only slightly for $x$ just below 1. All the curves for $n > 9$ look like a straight line corresponding to the value $s = 0.5$, which is the multiplier present in Schröder's equation (Eq. 8.). This also shows how exactly Eq. 8. is satisfied by the limit function defined by Eq. 9. and by the functions corresponding to relatively small values of $n$.

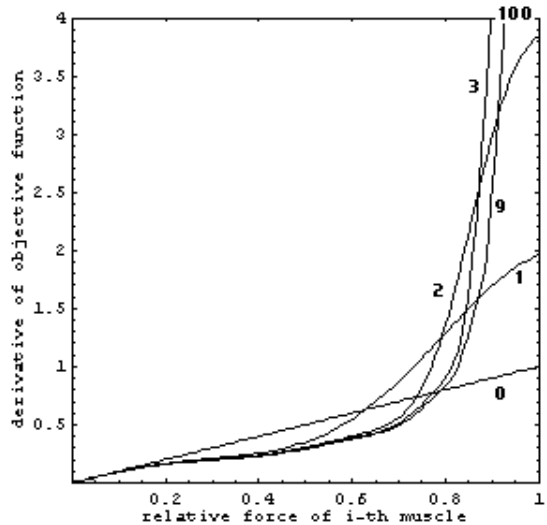

Fig. 3. A few first iterations according to formula (9); curves corresponding to n equal to 0, 1, 2, 3, 9 and 100 are shown

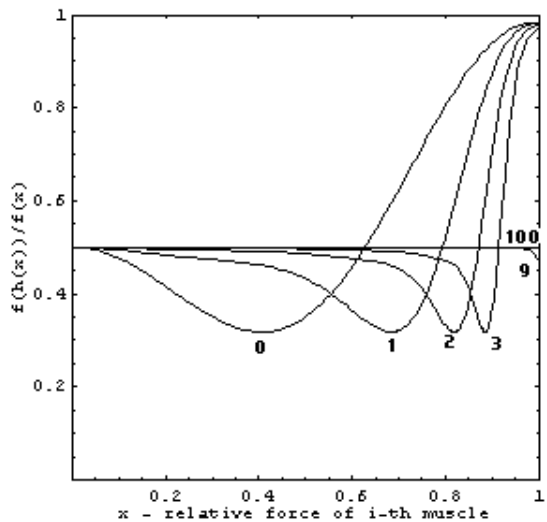

Fig. 4. Ratio $f(h(x))/f(x)$ as a function of $x$, the relative muscle force; curves corresponding to iteration numbers 0, 1, 2, 3, 9 and 100 are shown

The method proposed here to construct exactly the objective function whose minimization yields the observed force sharing patterns may be also used for other linearly constrained inverse optimization problems, e.g. those found in physics and economics. To the best of the

Can all muscular load sharing patterns be regarded as optimal in some sense?

author's knowledge, it makes it possible for the first time to deduce the very form of the objective function and not just a set of parameters in an earlier assumed function.